\documentclass[article,amsmath,amssymb,floatfix,aps]{revtex4}
\usepackage{graphicx}
\usepackage{epsfig}
\usepackage{multirow}
\usepackage{color}
\usepackage{cancel}

\begin{document}

\title{Parity Partner Bands in $^{163}$Lu: \\ A novel approach for describing the negative parity states from a triaxial super-deformed band}

\author{R.  Poenaru}%
 \email{robert. poenaru@drd. unibuc. ro}
 \affiliation{Doctoral School of Physics, University of Bucharest}%
 \affiliation{Horia Hulubei National Institute of Nuclear Physics and Engineering, Magurele}%
\author{A.  A.  Raduta}%
\email{raduta@nipne. ro}
\affiliation{Horia Hulubei National Institute of Nuclear Physics and Engineering, Magurele}%
\affiliation{Academy of Romanian Scientists, Bucharest}%

\date{\today}

\begin{abstract}
 The wobbling spectrum of $^{163}$Lu is described through a novel approach, starting from a triaxial rotor model within a semi-classical picture, and obtaining a new set of equations for all four rotational bands that have wobbling character.  Redefining the band structure in the present model is done by adopting the concepts of Signature Partner Bands and Parity Partner Bands.  Indeed, describing a wobbling spectrum in an even-odd nucleus through signature and parity quantum numbers is an inedited interpretation of the triaxial super-deformed  bands. 

\end{abstract}

\maketitle

 The wobbling motion was firstly described by Bohr and Mottelson within a particle triaxial rotor coupling, where the rotation axis moves on a curly cone.  This sort of motion is a signature of the triaxial nuclei, these being not much considered across the time. 
Although it was firstly predicted theoretically for even-even nuclei \cite{BMott}, this collective mode was also pointed out in several even-odd nuclei, with $^{163}$Lu being considered the best \emph{wobbler}, mainly due to its relatively rich spectrum: four triaxial super-deformed bands $TSD_{1,2,3,4}$.   The $TSD_1$ is interpreted as the ground state - yrast - band, while the other three as  wobbling multi-phonon excited bands \cite{odegaard2001evidence,Jens}. The common view on these bands is that the alignment of the odd-proton angular momentum, $i_{13/2}$ drives the system  to very large stable deformation \cite{frau}.  In the meantime, several neighboring odd-nuclei were identified as wobblers i. e., $^{161,165,167}$Lu \cite{Jens,Scho,Amro,Hage,Bring,Hage1}, and recently the nuclei $^{135}$Pr \cite{Matta,Sen}, $^{167}$Ta \cite{Bring,Hart}, $^{187}$Au \cite{Sen1}, $^{130}$Ba\cite{Chen}, $^{105}$Pd \cite{Timar}, 
$^{127}$Xe \cite{Chakr}, and $^{183}$Au \cite{Nand}.

In  a previous work \cite{raduta2020towards,raduta2020new}, a successful  description of the wobbling phenomenon in $^{163}$Lu was achieved.  Therein, the calculations were based on a particle-triaxial rotor system, that was semi-classically treated. 
The band structure was obtained in terms of two ground state bands ($TSD_1$ and $TSD_2$) of different signatures, given by  coupling  an odd $j=i_{13/2}$ proton to a core with angular momenta R=0,2,4,6,. .  and R=1,3,5,. . . , respectively, one wobbling phonon excitation  of the $TSD_2$ band, $n_w=1$, $TSD_3$, and one ground state band obtained by coupling  a different valence nucleon, namely the $j=h_{9/2}$ to a core exhibiting an angular momentum  from the sequence $\mathbf{R}=0,2,4,\dots$.  

Here we address the question whether the four TSD bands could be described by coupling a unique single particle state, i. e., $i_{13/2}$, to a core of a natural parity for $TSD_{1,2,3}$ and a core of negative parity and $\mathbf{R}=1,3,5,\dots$.  in the case of $TSD_4$.  Within this particle-core basis a similar Hamiltonian as in the previous paper is treated via a time dependent variational formalism.  In this manner one derives the classical equations of motion for the generalized canonical coordinates. 

For the sake of a self content presentation,in what follows we shall briefly introduce the necessary ingredients of the formalism.  

The Hamiltonian of $^{163}$Lu has a particle-rotor character and describes the interaction between an even-even triaxial core and a single nucleon that moves in the quadrupole deformed mean field generated by the core. 

\begin{align}
    H=H_\text{rot}+H_\text{sp}\ .  \label{hamiltonian_formula}
\end{align}

The first term represents the triaxial rotor Hamiltonian, with the core angular momentum  $\mathbf{R}=\mathbf{I}-\mathbf{j}$, and the inertial parameters $A_i$. 

\begin{align}
    H_\text{rot}=\sum_{i=1,2,3}A_i\left(I_i-j_i\right)^2,
\end{align}
 
 The inertial parameters $A_i$ are related to the moments of inertia (MoI)corresponding to the principal axes of the triaxial ellipsoid, through the equation $A_i=\frac{1}{2\mathcal{I}_i}$.  

The single-particle term from Eq.  \ref{hamiltonian_formula} is defined in terms of the triaxiality parameter $\gamma$ and the potential strength $V$. Actually  this term expresses the mean field for the single particle motion, determined by a collective quadrupole and a single particle quadrupole interaction \cite{Davyd}. 

\begin{align}
    H_\text{sp} = \frac{V}{j(j+1)}\left[\cos\gamma\left(3j_3^2-\mathbf{j}^2\right)-\sqrt{3}\sin\gamma\left(j_1^2-j_2^2\right)\right]+\epsilon_j.  \label{sp_hami}
\end{align}

The term $\epsilon_j$ from Eq.  \ref{sp_hami} represents the single particle energy. 
The eigenvalues of interest for $H$ are obtained on the base of a semi-classical approach. 
Thus, the total Hamiltonian $H$ is dequantized through  the time dependent variational equation (TDVE)\cite{rad16,rad17,Buda}:
\begin{equation}
\delta\int_{0}^{t}\langle \Psi_{IjM}|H-i\frac{\partial}{\partial t'}|\Psi_{IjM}\rangle d t'=0,
\end{equation}
where the trial function is chosen as:
\begin{equation}
|\Psi_{Ij;M}\rangle ={\bf N}e^{z I_-}e^{s j_-}|IMI\rangle |jj\rangle ,
\end{equation} 
with $I_-$ and ${j}_-$ denoting the lowering operators for the intrinsic angular momenta ${\bf I}$ and ${\bf j}$ respectively, while ${\bf N}$ is  the normalization factor. 
$|IMI\rangle $ and $|jj\rangle$ are extremal states for the operators ${\hat I}^2, {\hat I}_3$ and ${\hat j}^2, {\hat j}_3$, respectively.  We notice that the trial function is a mixture of  components of definite K, which is consistent with the fact that for triaxial nuclei, $K$ is not a good quantum number.  The name of TSD bands is the abbreviation for triaxial super-deformed bands suggesting that the ground band head state is an isomeric state with a relatively large half-life. 
 
The variables $z$ and $s$ are complex functions of time and play the role of classical phase space coordinates describing the motion of the core and the odd particle, respectively:
\begin{equation}
z=\rho e^{i\varphi},\;\;s=fe^{i\psi}. 
\end{equation}
Changing the variables $\rho$ and $f$ to  $ r$ and $t$, respectively:
\begin{equation}
r=\frac{2I}{1+\rho^2},\;\;0\le r\le 2I;\;\;
t=\frac{2j}{1+f^2},\;\; 0\le t\le 2j,
\end{equation}
the classical equations of motion acquire the canonical Hamilton form:
\begin{equation}
\frac{\partial {\cal H}}{\partial r}=\stackrel{\bullet}{\varphi},\;\frac{\partial {\cal H}}{\partial \varphi}=-\stackrel{\bullet}{r};\;
\frac{\partial {\cal H}}{\partial t}=\stackrel{\bullet}{\psi};\;\frac{\partial {\cal H}}{\partial \psi}=-\stackrel{\bullet}{t}.  
\label{eqmot}
\end{equation}
where ${\cal H}$ denotes the average of $H$ with the trial function $|\Psi_{IjM}\rangle$ and plays the role of the classical energy function. 
The classical energy has the expression : 
\begin{equation}
{\cal H}(r,\varphi;t,\psi)=\langle \Psi_{IjM}|H|\Psi_{IjM}\rangle\nonumber\\
\label{classen}
\end{equation} 
and is minimal (${\cal H}^{(I,j)}_{min}$) in the point
$(\varphi,r)=(0,I);(\psi,t)=(0,j)$, when $A_1<A_2<A_3$.  
The equations of motion provided by the variational principle are:
\begin{eqnarray}
\stackrel{\bullet}{\varphi}&=&\frac{2I-1}{I}(I-r)\left(A_1\cos^2\varphi+A_2\sin^2\varphi-A_3\right)\nonumber\\
                           &-&2\sqrt{\frac{t(2j-t)}{r(2I-r)}}(I-r)\left(A_1\cos\varphi\cos\psi+A_2\sin\varphi\sin\psi\right)+2A_3(j-t),\nonumber\\
\stackrel{\bullet}{\psi}&=&\frac{2j-1}{j}(j-t)\left(A_1\cos^2\psi+A_2\sin^2\psi-A_3\right)\nonumber\\
                           &-&2\sqrt{\frac{r(2I-r)}{t(2j-t)}}(j-t)\left(A_1\cos\varphi\cos\psi+A_2\sin\varphi\sin\psi\right)+2A_3(I-r)\nonumber\\
                           &-&V\frac{2j-1}{j^2(j+1)}(j-t)\sqrt{3}\left(\sqrt{3}\cos\gamma+\sin\gamma\cos2\psi\right),\nonumber\\
-\stackrel{\bullet}{r}&=&\frac{2I-1}{2I}r(2I-r)\left(A_2-A_1\right)\sin2\varphi\nonumber\\
                           &+&2\sqrt{t(2j-t)r(2I-r)}\left(A_1\sin\varphi\cos\psi-A_2\cos\varphi\sin\psi\right),\nonumber\\
-\stackrel{\bullet}{t}&=&\frac{2j-1}{2j}t(2j-t)\left(A_2-A_1\right)\sin2\psi\nonumber\\
                      &+&2\sqrt{r(2I-r)t(2j-t)}\left(A_1\cos\varphi\sin\psi-A_2\sin\varphi\cos\psi\right)\nonumber\\
                           &+&V\frac{2j-1}{j^2(j+1)}t(2j-t)\sqrt{3}\sin\gamma\sin2\psi .
\end{eqnarray}
Linearizing the equations of motion around the minimum point of ${\cal H}$, one obtains a harmonic motion for the system, with the frequency given by the equation:
\begin{equation}
\Omega^4+B\Omega^2+C=0,
\label{ecOm}
\end{equation}
where the coefficients B and C have the expressions (10) and (11) from Ref \cite{raduta2020new}.

Under certain restrictions for MoI's the dispersion equation (\ref{ecOm}) admits two real and positive solutions, which here after  will be denoted by $\Omega^{I}_1$ and $\Omega^{I}_{2}$ for $j=i_{13/2}$and ordered as: $\Omega^I_1<\Omega^I_{2}$ . 

Further, to the $TSD_{1,2,3,4}$ bands we associate the energies:
\begin{eqnarray}
  &&  E_I^\text{TSD1}=\epsilon_{j} + \mathcal{H}_\text{min}^{(I,j)}+\mathcal{F}_{00}^I , \;\;I=R+j, R=0,2,4,. . . ,\nonumber \\
  &&  E_I^\text{TSD2}=\epsilon_{j,1} + \mathcal{H}_\text{min}^{(I,j)}+\mathcal{F}_{00}^I , \;\;I=R+j, R=1,3,5,. . . \nonumber\\
  &&  E_I^\text{TSD3}=\epsilon_{j} + \mathcal{H}_\text{min}^{(I,j)}+\mathcal{F}_{10}^I, \;\; I=R+j, R=0,2,4,. . .  \nonumber\\
  &&  E_I^\text{TSD4}=\epsilon_{j,2} + \mathcal{H}_\text{min}^{(I,j)}+\mathcal{F}_{00}^I,\;\;I=R+j,\;R=1,3,5,. . . . \nonumber\\ \label{wobbling_energies}
\end{eqnarray}
where $\mathcal{F}_{n_{w_1}n_{w_2}}$ is function of the wobbling frequencies

\begin{align}
    F_{n_{w_1}n_{w_1}}^I=(n_{w_1}+\frac{1}{2})\Omega_1^I+(n_{w_2}+\frac{1}{2})\Omega_2^I.  \label{phonons}
\end{align}
while $\mathcal{H}^{(I,j)}_{min}$  is the minimal classical energy.  We considered different re-normalizations for the single-particle mean field in the signature unfavored as well as in the negative parity states, which result two distinct energy shifts for the excitation energies in the TSD2 and TSD4 bands, respectively. These two quantities will be adjusted throughout the numerical calculations such that the energy spectrum is best reproduced.  The phonon numbers corresponding to the four bands are listed in Table \ref{tabular_phonon_numbers}, where values of the parity and signatures are also shown. 
In a previous publication \cite{raduta2020new} one showed that the signature is a good quantum number. 
One can  prove that parity is also a good quantum number in our formalism.  Indeed, taking into account that the parity operator is a product of the complex conjugation operation and a rotation of angle $\pi$ around the 2-axis ($P=e^{-i\pi J_2}C$) and acting on the trial function with the total parity operator $P_t=P_cP_{sp}$ one obtains:
\begin{equation}
P_t\Psi(r,\varphi;t,\psi)=\Psi(r,\varphi+\pi; t, \psi +\pi). 
\end{equation}
On the other hand the energy function is invariant at changing the angles with $\pi$:
\begin{equation}
{\cal{H}}(r,\varphi+\pi; t, \psi +\pi)={\cal{H}}(r,\varphi;t,\psi). 
\end{equation}
This induces the fact that the functions $\Psi$ and its image through $P_t$ are linearly dependent, differing by a multiplicative constant of modulus equal to unity.  Thus,
\begin{equation}
\Psi(r,\varphi+\pi; t, \psi +\pi)=\pm \Psi(r,\varphi;t,\psi). 
\end{equation}
The above result is a reflection of the fact that the triaxial rotor admits eigenfunctions of negative parity.  Indeed, let $r_k$ ,k=0,1,2,3 be the eigenvalues of the four elements of the group $D_2$:
${\cal{E}},e^{-i\pi R_1}, e^{-i\pi R_2}, e^{-i\pi R_3}$ with ${\cal{E}}$ denoting the unity rotation.  The eigenfunctions of the rotor Hamiltonian being at a time  eigenfunctions for the $D_2$ elements form irreducible representation of the group, with the eigenvalues $(r_0,r_1,r_2,r_3)$.  Two of these irrers have negative parity.  These are: $(1, -1, -1, 1)$ and $(1, 1, -1, -1)$. 

 The spin sequences for the TSD bands are shown in Table \ref{spin_sequences}. 

\begin{table}[h]
    \centering
  \begin{tabular}{lllll}
  \hline
Band & $n_{w_1}$ & $n_{w_2}$ &  $\pi$ &  $\alpha$ \\
\hline
\hline
$TSD_1$ &     0      &       0    &    +1  &    +1/2  \\
$TSD_2$ &    0       &       0    &    +1  &    -1/2  \\
$TSD_3$ &     1      &     0      &    +1  &    +1/2  \\
$TSD_4$ &     0      &     0      &     -1  &    -1/2  \\
\hline
\end{tabular}
    \caption{The wobbling phonon numbers, parities and signatures assigned for the triaxial bands in $^{163}$Lu  within the model. }
    \label{tabular_phonon_numbers}
\end{table}

\begin{table}[h]
    \centering
  \begin{tabular}{llll}
  \hline
Band & j & $\mathbf{R}$-sequence & $\mathbf{I}$-sequence \\
\hline
\hline
$TSD_1$ & $i_{13/2}$  &   $0,2,4,\dots$         &   $13/2,17/2,21/2,\dots$         \\
$TSD_2$ & $i_{13/2}$  &   $1,3,5,\dots$         &           $27/2,31/2,35/2,\dots$ \\
$TSD_3$ & $i_{13/2}$  &  $0,2,4,\dots$ & $33/2,37/2,41/2,\dots$     \\
$TSD_4$ & $i_{13/2}$  &         $1,3,5,\dots$   &       $47/2,51/2,55/2,\dots$    
\end{tabular}
    \caption{The spin sequences that belong to the wobbling spectrum of $^{163}$Lu, where $j$ is the $i_{13/2}$-odd proton.  }
    \label{spin_sequences}
\end{table}

In what follows it is worth  analyzing  the dependence of the classical energy function on the Cartesian coordinates $x_k=I_k,\;k=1,2,3$:

\begin{equation}
    x_1=I\sin\theta \cos\varphi\ ,\;\;
    x_2=I\sin\theta \sin\varphi\ ,\;\;
    x_3=I\cos\theta. 
\end{equation}

In polar coordinates the classical energy function reads:

\begin{eqnarray}
    \mathcal{H}&=&I\left(I-\frac{1}{2}\right)\sin^2\theta\left(A_1\cos^2\varphi+A_2\sin^2\varphi-A_3\right)-\nonumber\\
    &-&2A_1Ij\sin\theta+T_{rot}+T_{sp}\ , \label{energy_function}
\end{eqnarray}
\noindent
where the last two terms are independent of the coordinates and have the forms:

\begin{align}
    T_{rot}&=\frac{I}{2}(A_1+A_2)+A_3I^2\ ,\\
    T_{sp}&=\frac{j}{2}(A_2+A_3)+A_1j^2-V\frac{2j-1}{j+1}\sin\left(\gamma+\frac{\pi}{6}\right)\ . 
\end{align}
In obtaining this expression the single particle terms were considered in the minimum point. 

The classical energy admits two constants of motion: the system energy and the total angular momentum.  Therefore, the classical trajectories are determined by intersecting the surfaces describing the two constants of motion, that are an ellipsoid and a sphere, respectively:

\begin{align}   
    &E=T_{rot}+T_{sp}-I\left(I-\frac{1}{2}\right)A_3-2A_1Ij+\left(1-\frac{1}{2I}\right)A_1x_1^2\nonumber\\
&+\left(1-\frac{1}{2I}\right)A_2x_2^2
      +\left[\left(1-\frac{1}{2I}\right)A_3+A_1\frac{j}{I}\right]x_3^2, \nonumber \\
       \label{ellipsoid_rotation}\nonumber \\
  &I^2=x_1^2+x_2^2+x_3^3. 
\end{align}

 For a given total angular momentum and a given set of MoI's one can solve the above equations by expressing two unknowns in term of the third one and thus, classical trajectories of an wobbling character are obtained. 

\begin{figure}
    \centering
    \includegraphics[scale=0.50]{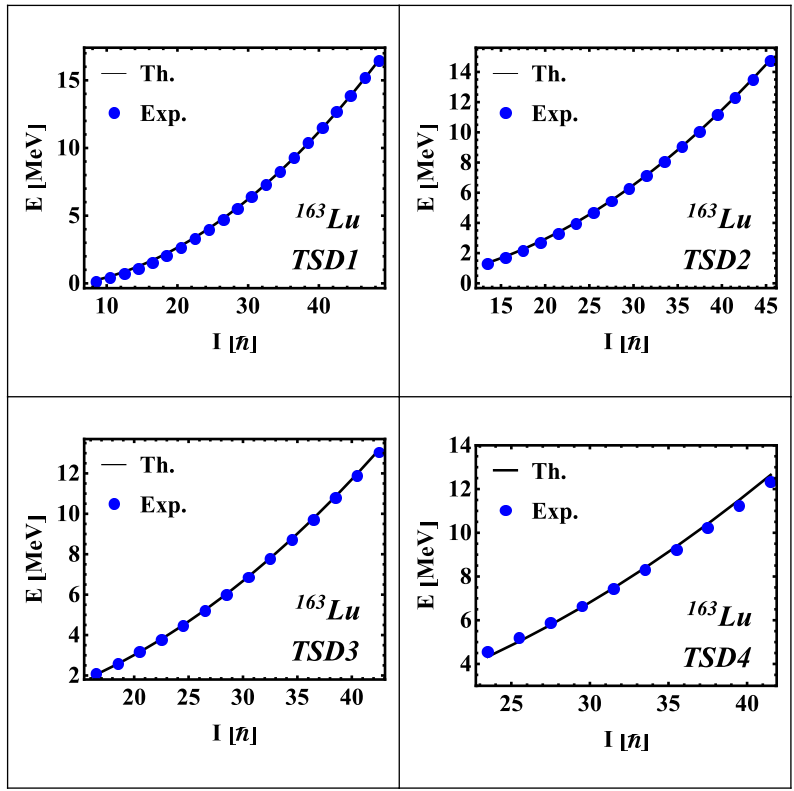}
    \caption{The excitation energies for the bands $TSD_1$, $TSD_2$, $TSD_3$, and $TSD_4$. }
    \label{tsd_bands}
\end{figure}

We  now  proceed at discussing the numerical results.  As already mentioned  the application is made for $^{163}$Lu, since this is the only isotope exhibiting both positive and negative parity bands
and thus one can check the validity of our proposed formalism. 
 By using the expressions \ref{wobbling_energies}, a least squares fitting procedure was used for finding the parameter set $\mathcal{P}=(\mathcal{I}_1,\mathcal{I}_2,\mathcal{I}_3,\gamma,V)$.  The found values of $\mathcal{P}$, are  shown in Table \ref{parameter_set},leading to an r.m.s. value of $\approx 79$ keV, which is much better than that obtained through a different approach
 \cite{raduta2020new}, where this  is $\approx 240$ keV.  Keep in mind that the fitting procedure was done simultaneously for all four bands,contrary to Ref. \cite{raduta2020new} where a separate parameter set for TSD4 was used, invoking a different polarization effect of the core, due to the particle-core interaction.  Concerning the single particle energies, the unfavored signature as well as the negative parity states induce a correction for the mean field with the quantities: $\epsilon_{j,1}-\epsilon_{j}=0. 3$ MeV and $\epsilon_{j,2}-\epsilon_{j}=0. 6$ MeV, respectively.  Results of our calculations are compared with the corresponding data in Fig.  \ref{tsd_bands}, where one notices a very good agreement of the two sets of energies. 

The quantity $\epsilon_{j,1}-\epsilon_{j}$represents the contribution of the single particle mean field to the signature splitting. This is added to the splitting due to the core, caused by the fact that two distinct TDVE's were used for the two partner bands which results in having an I-dependence of the $\cal{H}$, $\Omega_1$ and $\Omega_2$.
The total signature splitting for the head and terminus states of $TSD_2$ are $E(27/2)-E(25/2)=492$ keV and $E(91/2)-E(89/2)=936$ keV, respectively, which agrees with the estimate of Ref. \cite{Jens}. Note that within microscopic approaches the signature splitting is determined by a deformed single particle basis plus a cranking constrain, while here by  applying the TDVE for each angular momentum and the corresponding correction due to the single particle energies.  

Remark the fact that the difference $E_I^\text{TSD4}-E_I^\text{TSD2}\approx 300 keV$, which suggests that the states of the same angular momentum  from the TSD2 and TSD4 bands might emerge through the parity projection from a sole function without space reflection symmetry.  In our case, this is caused by the fact that the wobbling frequency is parity independent. These bands are thus parity partners, as defined in Refs.  \cite{chas,Rad1,Rad2,Rad3}. From the above arguments it results that within the angular momentum space the action of the parity operator to any angular momentum vector {\bf I} leads to 
-${\bf I}$. Consequently, the parity operator commutes with the quantal Hamiltonian $H$ and therefore the eigenfunctions of $H$ are of either positive or  negative parity. Moreover, the states of different parities are degenerate. In order to lift up this degeneracy, an additional term linear in the total angular momentum is to be included in $H$. Since such a term is missing the ad-hoc correction of the mean-field with the amount of 0.6 MeV for the $TSD_4$ states is necessary. In this way one simulates the breaking of the parity symmetry. By contrast, within a microscopic formalism  one starts with a single particle basis generated by a mean field without space reflection symmetry. By different procedures one finally finds the many body wave function which is a mixture of both parities.
Restoring the parity symmetry is achieved by selecting from the wave function only the components of a definite parity, i.e., projecting out the good  parity, which results in having a doublet structure of positive and negative parity states in the spectrum of $H$. Effects of the stable octupole deformation on the rotational motion in nuclei was investigated in Ref.\cite{Naz}.

In Ref.\cite{Jens1} the band $TSD_4$ was interpreted  as having a three quasiparticle state structure, although it is not yet proved that such an interpretation is unique. Our new view on the $TSD_4$ band is based on the simple assumption that the negative parity states of the core play an important role. The very good agreement with the data pleads in favor of a realistic description.
Note that to a certain extent our interpretation is consistent with that of Ref.\cite{Jens1}. Indeed, the core of $^{163}$Lu cosists of Z=70 protons and N=92 neutrons, the system being of positive parity in its ground state. If one promotes a proton from $i_{13/2}$ to an orbital like $f_{7/2}$ or $h_{9/2}$, then the many body system passes to an excited state of negative parity.
Thus, in our case the core turns out to be an effective core of negative parity involving the 2qp excitations. Adding the odd $i_{13/2}$  proton to the core one has the third qp excitation. Concluding the hypothesis of a triaxial rotor core of negative parity simulates the  mentioned microscopic picture.

\begin{table}[h]
    \centering
  \begin{tabular}{lllll}
  \hline
$\mathcal{I}_1$ [$\hbar^2$/MeV] & $\mathcal{I}_2$ [$\hbar^2$/MeV]& $\mathcal{I}_3$ [$\hbar^2$/MeV] & $\gamma$ [deg. ] & $V$ [MeV] \\
\hline
\hline
72              & 15              & 7               & 22       & 2.1
\end{tabular}
    \caption{The parameter set $\mathcal{P}$ that was determined by a fitting procedure of the excitation energies of $^{163}$Lu. }
    \label{parameter_set}
\end{table}

\begin{figure}
    \centering
    \includegraphics[scale=0.5]{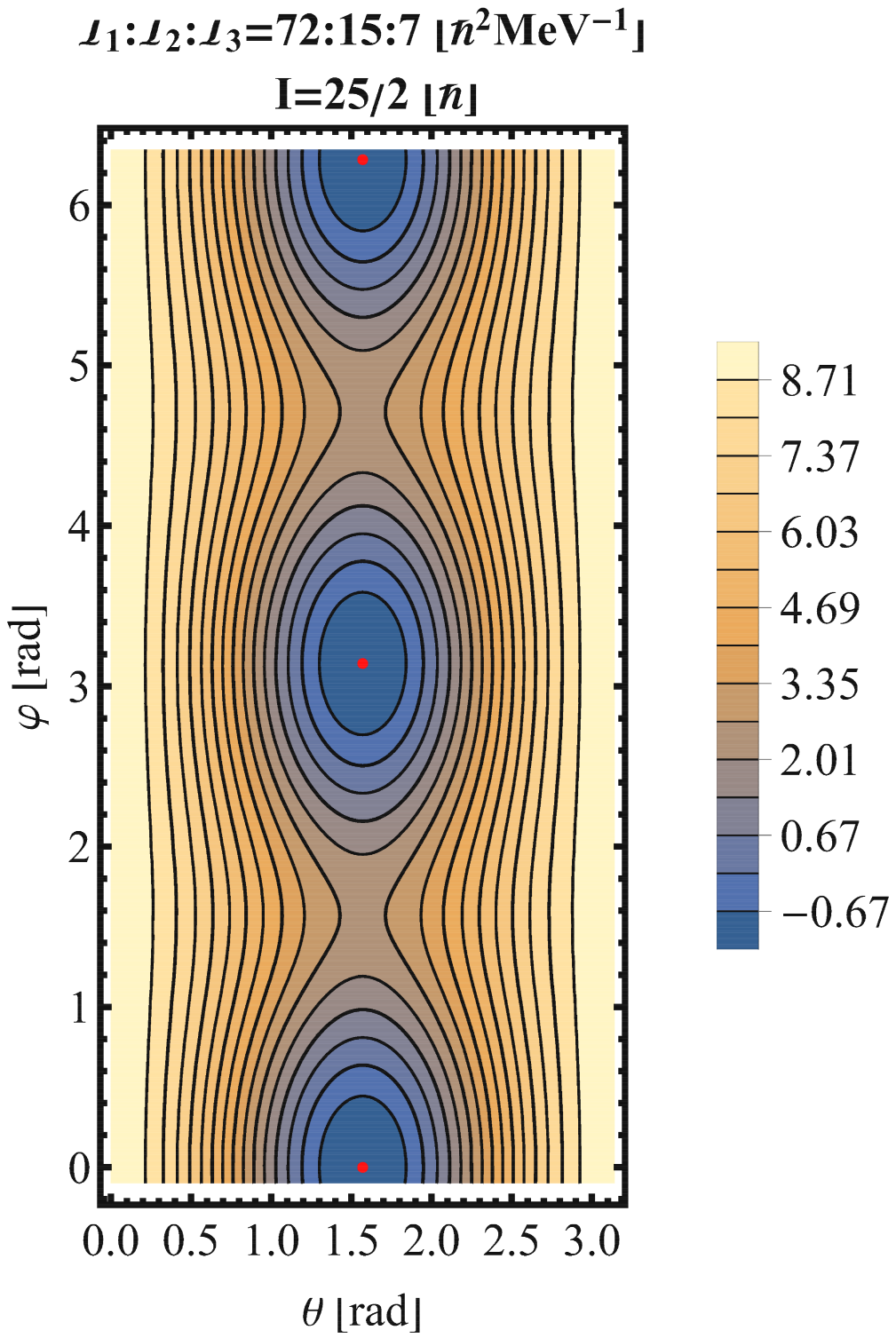}
    \includegraphics[scale=0.5]{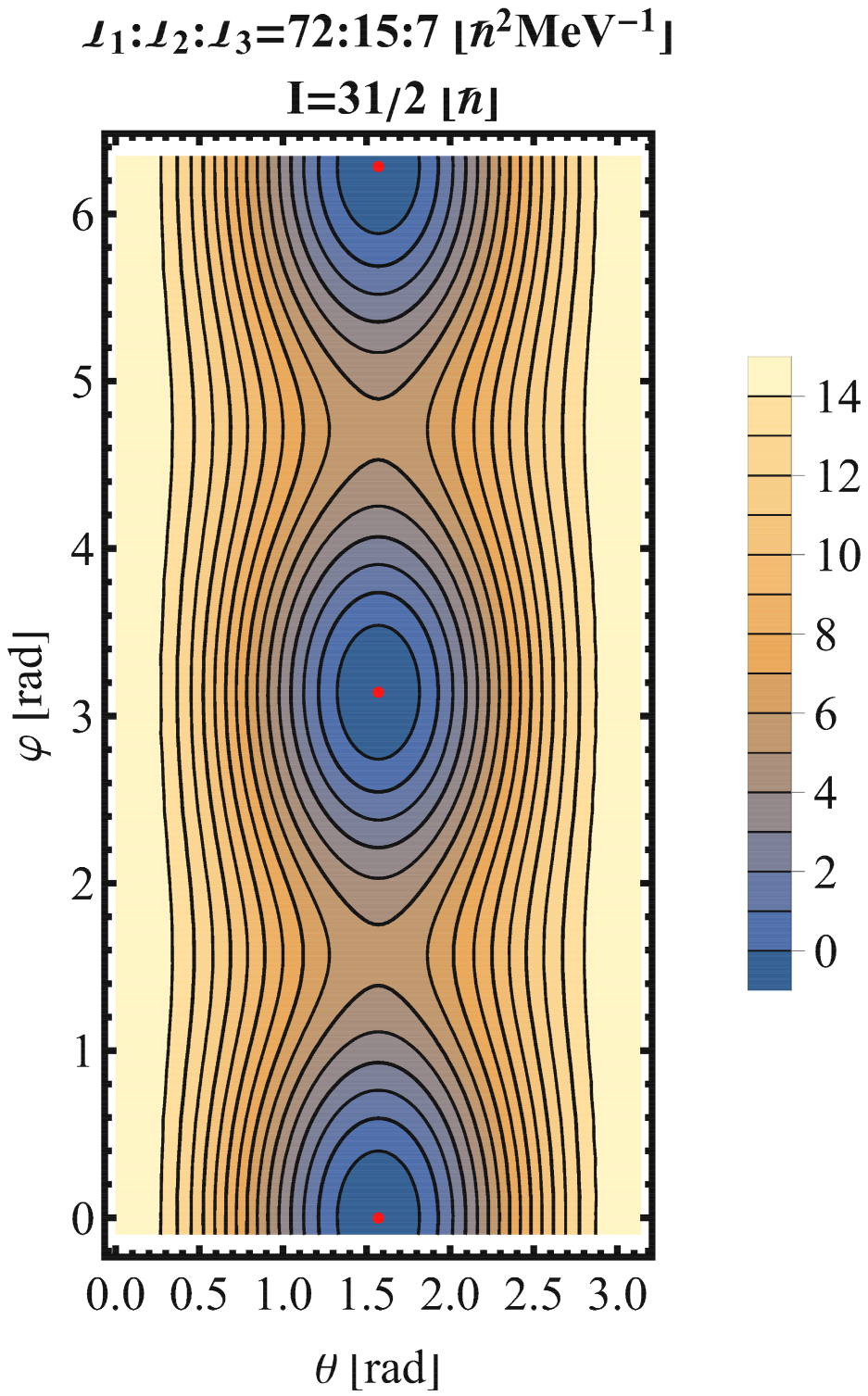}
    \caption{A contour plot with the energy function $\mathcal{H}$ for $TSD_1$ and $TSD_2$.  The parameter set $\mathcal{P}$ was used for the numerical calculations. }
    \label{contour-tsd1}
\end{figure}

\begin{figure}
    \centering
    \includegraphics[scale=0.5]{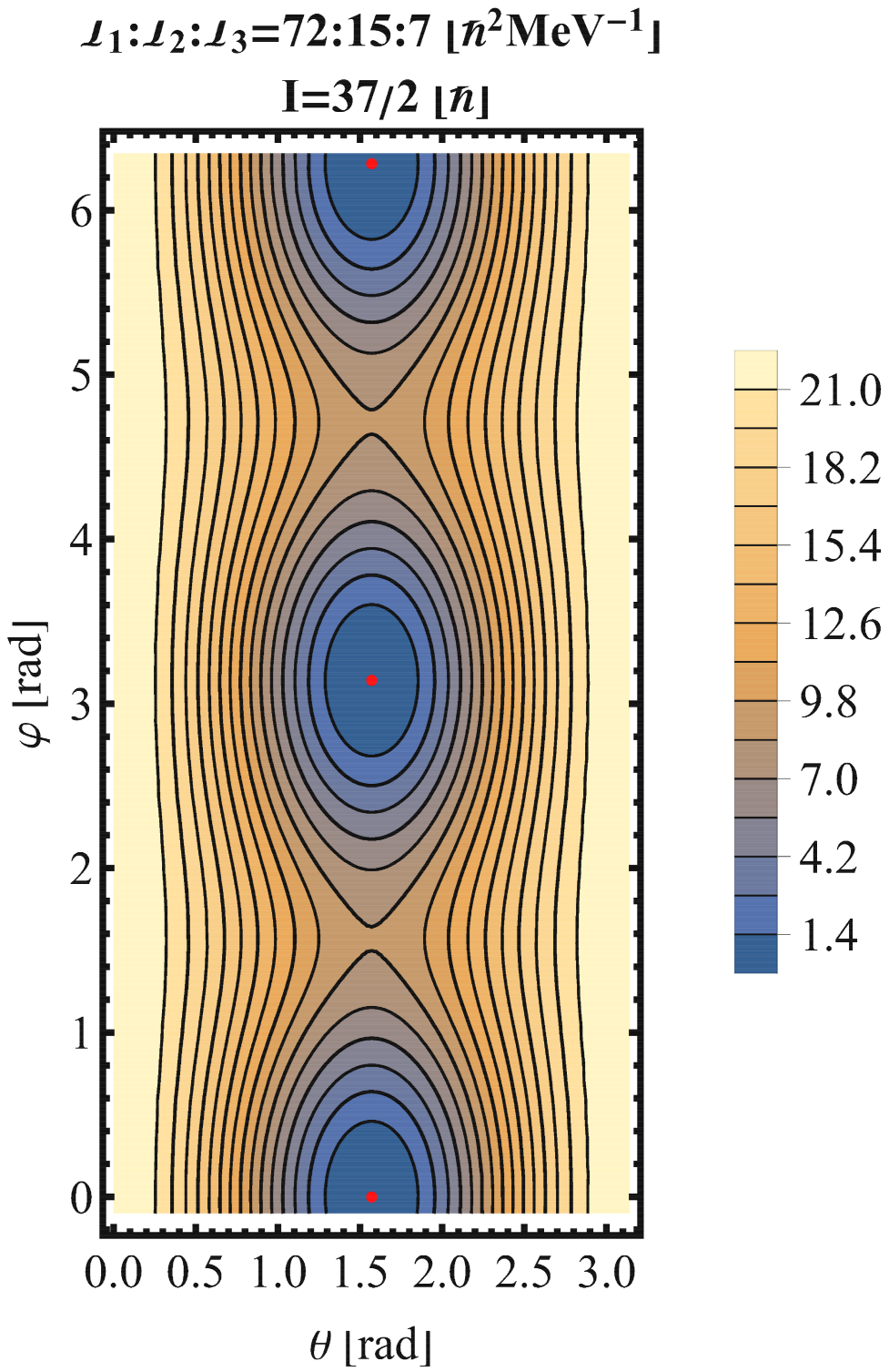}
    \includegraphics[scale=0.5]{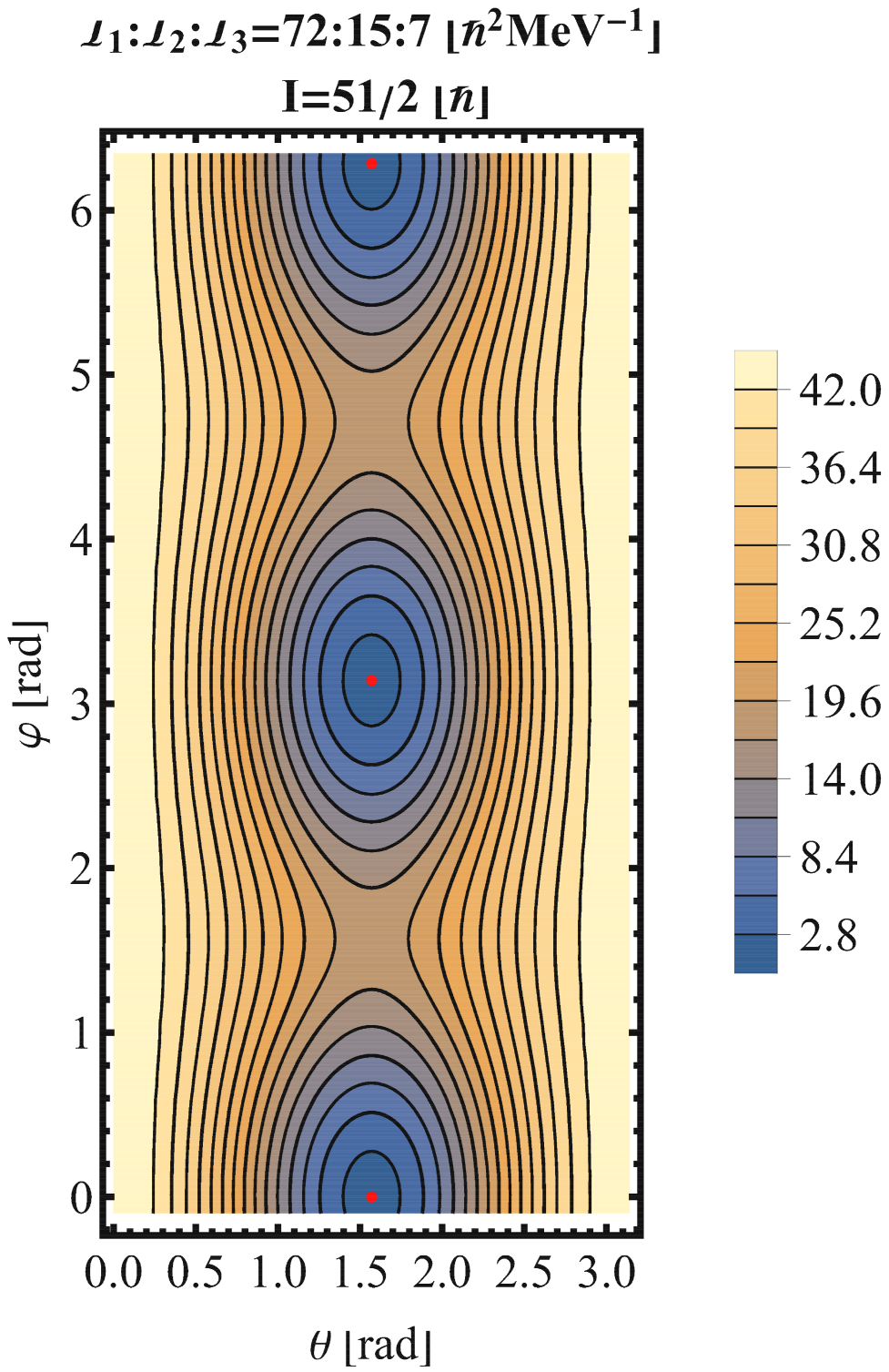}
    \caption{A contour plot with the energy function $\mathcal{H}$ for $TSD_3$ and $TSD_4$.  The parameter set $\mathcal{P}$ was used for the numerical calculations. }
    \label{contour-tsd3}
\end{figure}

In terms of the stability of the wobbling motion with respect to the total angular momentum, several contour plots were plotted, using the obtained parameter set $\mathcal{P}$ with the help of Eq.  \ref{energy_function}.  For each band, a spin close to the band head of each sequence was chosen.  Due to the obtained MOI ordering, the surfaces have minimum points indicated by the red dots for each figure.  Results can be seen in Figs.  \ref{contour-tsd1},\ref{contour-tsd3}.  The four figures have many similarities suggesting common collective properties, but also differences caused by the fact that minima have different depths. The common feature consists of that the equi-energy curves surround a sole minimum for low energy while for higher energies the trajectories go around all minima, the lack of localization indicating an unstable picture. 

Finally we are interested in finding out the dependence of the classical trajectories on angular momenta as well as on energies.  Indeed, when the model Hamiltonian is diagonalized for a given I,  a set of $2I+1$ energies are obtained.  Therefore, it makes sense to study the trajectory change at increasing the energy. Trajectories are represented as the manifold given by intersecting the surfaces corresponding to the two constants of motion.  The first energy in each row corresponds to the real excitation energy  for that particular spin state, the second one represents the point at which the ellipsoid touches the sphere at the equator, which marks a nuclear phase transition - while the third one is the trajectory of the system at energies sufficiently large that the system changes its wobbling regime.  For low energies, one notices two distinct trajectories having as rotation axes the 1-axis and -1-axis, respectively.  As energy increases the two trajectories approach each other which results a tilted rotation axis for each of trajectories, the rotation axes being dis-aligned.  Note that this picture is fully consistent with that of Ref. \cite{Lawr}. When the two trajectories intersect each other, the trajectories surround both minima.  Increasing the energy even more one arrives again at two trajectories regime but with different rotation axes which become close to the 3-axis.  This reflects another phase transition for the system. 

\begin{figure}
    \centering
    \includegraphics[scale=0.55]{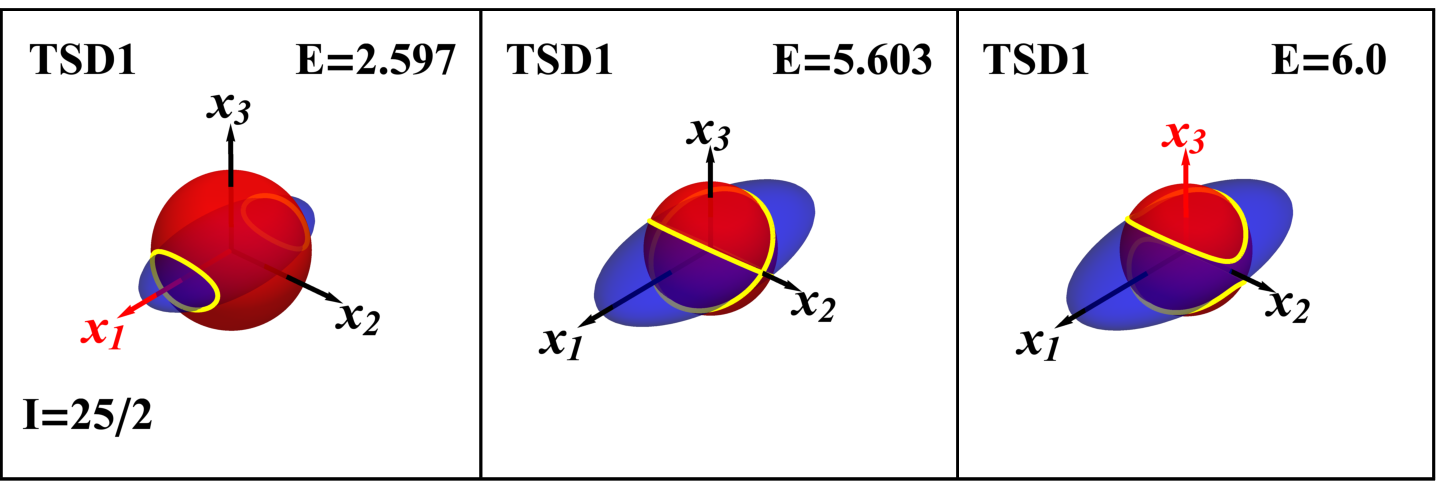}\\
    \includegraphics[scale=0.55]{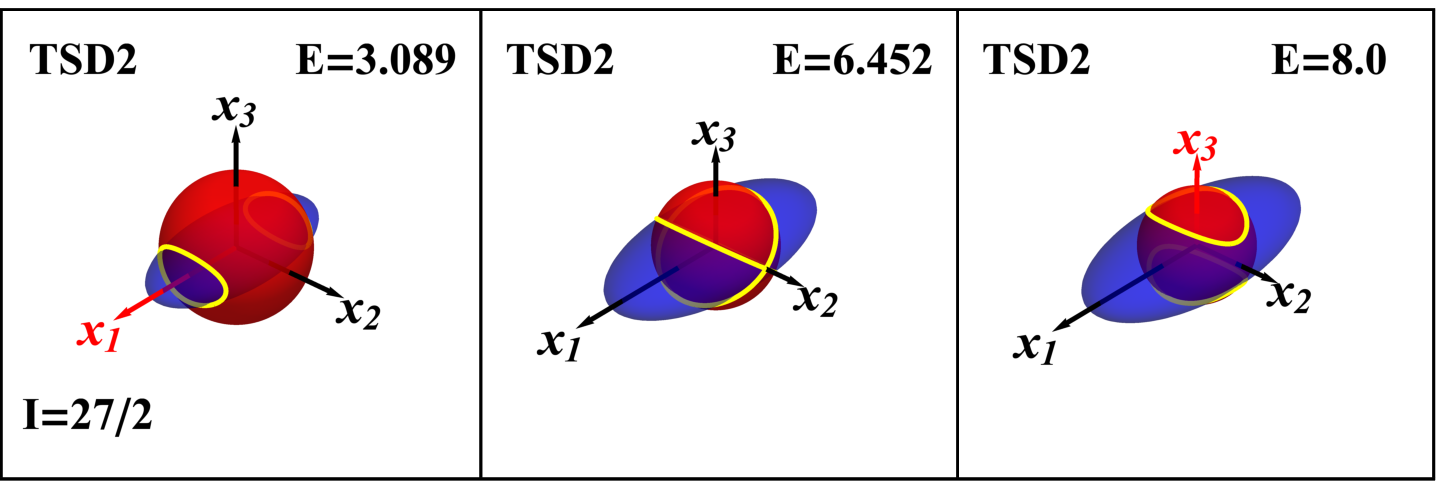}\\
    \includegraphics[scale=0.55]{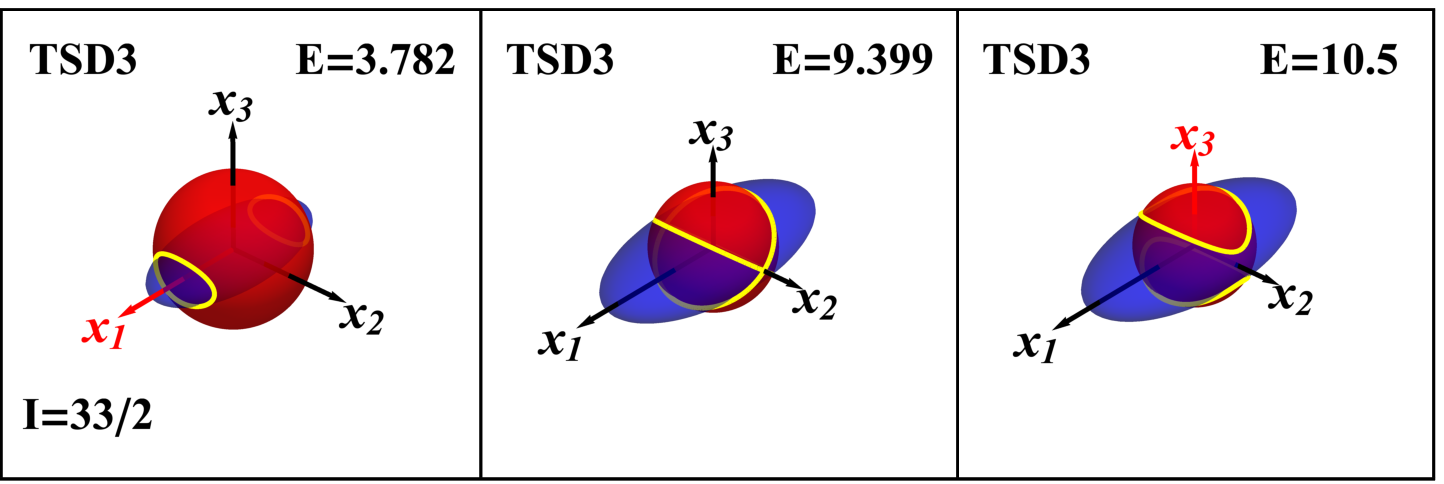}\\
    \includegraphics[scale=0.55]{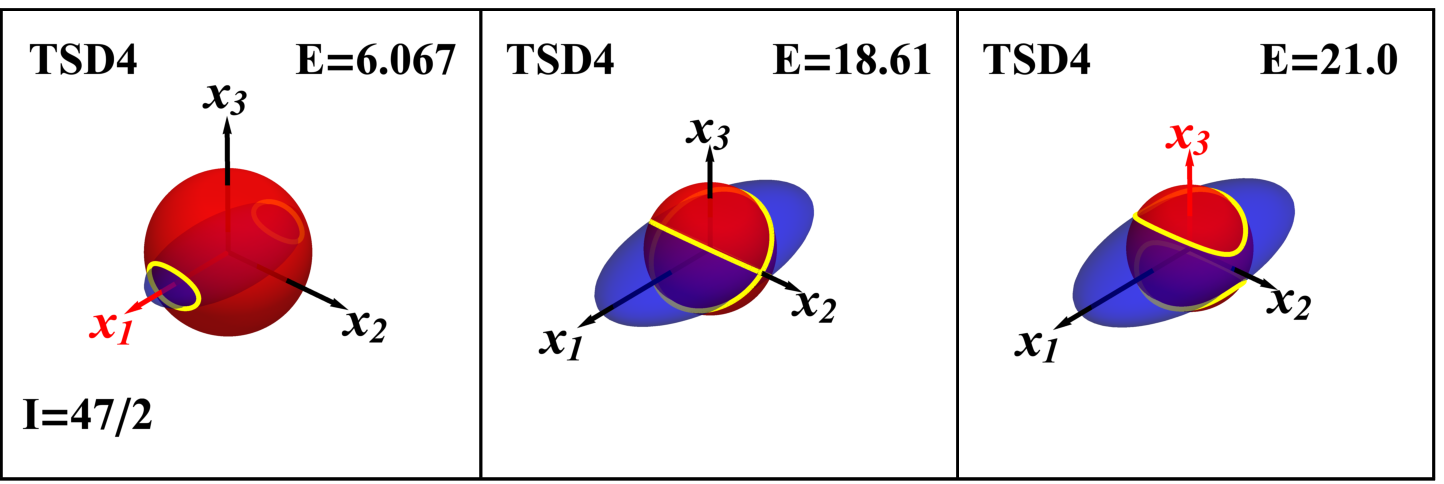}
    \caption{The nuclear trajectory of the system for a spin state belonging to each of the four TSD bands of $^{163}$Lu.  Intersection line marked with yellow color represents the actual orbits. }
    \label{ellipsoids-tsd1}
\end{figure}

The results of our investigation can be summarized as follows.  Despite the fact that $TSD_4$ is of an opposite parity than the lower bands, the four bands are described by coupling a sole single particle of positive parity to the core states of positive parity for $TSD_{1,2,3}$ and negative parity for $TSD_4$.  The core is not changing, which results in having a unique set of MoI-s but the mean field for the valence nucleon is modified for unfavored signature as well as for the negative parity band.  The contour plots for one representative state from each band, shows a similar structure but different depths and reaching the unstable regimes at different energies.  The system's trajectories corresponding to the four bands, obtained by intersecting the surfaces associated to the two constants of motion, the energy and the angular momentum squared , indicate that for low energy the rotation axes are the 1-axis and -1-axis defining two disjoint trajectories, while for higher energy the rotation axes are tilted toward the 3-axis. There are signals that $TSD_2$ and $TSD_4$ are parity partner bands.  Likewise the bands $TSD_1$ and $TSD_2$ are signature partner bands.  The electromagnetic  properties of these bands have been successfully described in Ref. \cite{raduta2020new}.  Obviously, the results from the quoted paper are valid also here. 

Concluding, the present model is a successful tool for accurately describing the wobbling spectrum of $^{163}$Lu, but also for understanding the rotational motion of the nuclear system with respect to its total spin. 

{\bf Acknowledgments}. This work was supported by UEFISCU through the project PCE-PN3-2021/0149

  
\end{document}